\documentstyle[12pt,epsf]{article}
\begin{document}

\thispagestyle{empty}
\begin{flushright}
MPI-Ph/96-59\\
TUM-T31-99/96\\
FERMILAB-PUB-96/191-T\\
hep-ph/9607447\\
July 1996
\end{flushright}
\vspace*{2cm}
\centerline{\Large\bf $K\to\pi\nu\bar\nu$ and High Precision Determinations}
\centerline{\Large\bf of the CKM Matrix}
\vspace*{1.5cm}
\centerline{{\sc Gerhard Buchalla$^3$} 
and {\sc Andrzej J. Buras$^{1,2}$}}
\bigskip
\centerline{\sl $^1$Physik Department, Technische Universit\"at
                M\"unchen}
\centerline{\sl D-85748 Garching, Germany}
\vskip0.6truecm
\centerline{\sl $^2$Max-Planck-Institut f\"ur Physik--
                Werner-Heisenberg-Institut}
\centerline{\sl F\"ohringer Ring 6, D-80805 M\"unchen, Germany} 
\vskip0.6truecm
\centerline{\sl $^3$Theoretical Physics Department}
\centerline{\sl Fermi National Accelerator Laboratory}
\centerline{\sl P.O. Box 500, Batavia, IL 60510, U.S.A.}

\vspace*{1.5cm}
\centerline{\bf Abstract}
\vspace*{0.2cm}
\noindent 
We investigate the future determination of the CKM matrix
using theoretically clean quantities, 
like $B(K^+\to\pi^+\nu\bar\nu)$,
$B(K_L\to\pi^0\nu\bar\nu)$ or $\sin 2\beta$, $\sin 2\alpha$ as
extracted from CP violation studies in $B$ physics.
The theoretical status of $K\to\pi\nu\bar\nu$ is briefly reviewed
and their phenomenological potential is compared with that of
CP asymmetries in $B$ decays. We stress the unique opportunities
provided by measuring the CP violating rare decay 
$K_L\to\pi^0\nu\bar\nu$. It is pointed out that this mode is
likely to offer the most precise determination of
${\rm Im}V^*_{ts}V_{td}$ and the Jarlskog parameter $J_{CP}$, the
invariant measure of CP violation in the Standard Model.
\vspace*{0.5cm}
\noindent 

\vspace*{1.2cm}
\noindent
PACS numbers: 12.15.Hh, 13.20.Eb 

\vfill

\newpage
\pagenumbering{arabic}

\section{Introduction}
\label{intro}

\begin{figure}[ht]
   \vspace{0cm}
   \epsfxsize=10cm
   \centerline{\epsffile{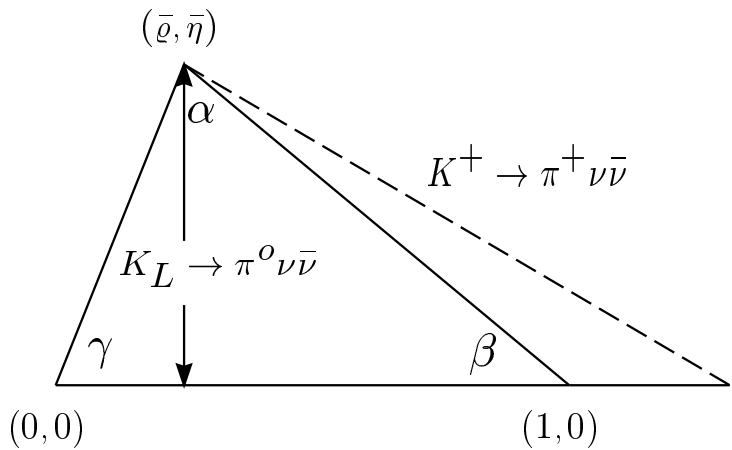}}
   \vspace*{-0cm}
\caption{\label{utfig} The unitarity triangle.}
\end{figure}

The Standard Model (SM) provides an economical and elegant
description of CP violation. Within the 
Cabibbo-Kobayashi-Maskawa (CKM) framework \cite{KM} the violation of
CP symmetry is accounted for by a single phase, naturally
emerging in the three generation model, and CP violation is
intimately connected to the physics of quark mixing. Until today
this theoretical ansatz is consistent with all known weak decay
phenomena, but some of the CKM parameters are only rather loosely
constrained and the information on CP violation is limited to
the $K^0-\bar K^0$ system.
\\
One of the most important goals of particle physics in the coming
years will be to precisely determine all parameters of the CKM
matrix and to check the SM picture for consistency by using as
many independent observables as possible. 
\\
In the standard parametrization of the CKM matrix \cite{PDG}
the four basic parameters are $s_{12}$, $s_{23}$, $s_{13}$ and
the phase $\delta$. A convenient alternative representation
uses the Wolfenstein parameters $\lambda$, $A$, $\varrho$ and
$\eta$ \cite{WO}, which can be defined by \cite{BLO,BBL}
\begin{equation}\label{wodef}
s_{12}=\lambda\qquad s_{23}=A\lambda^2\qquad
s_{13} e^{-i\delta}= A\lambda^3 (\varrho-i\eta)
\end{equation}
Here $|s_{13}|=|V_{ub}|$ and, to an accuracy of ${\cal O}(10^{-5})$,
$s_{12}=V_{us}$, $s_{23}=V_{cb}$. In the Wolfenstein
parametrization $\lambda=0.22$ can be used as an expansion
parameter to simplify expressions for CKM elements. The 
representation is particularly convenient for the unitarity triangle,
which graphically displays the unitarity relation
\begin{equation}\label{utdef}
1+\frac{V_{td}V^*_{tb}}{V_{cd}V^*_{cb}}=
-\frac{V_{ud}V^*_{ub}}{V_{cd}V^*_{cb}}\equiv \bar\varrho+i\bar\eta
\end{equation}
in the $(\bar\varrho, \bar\eta)$ plane (Fig. \ref{utfig}).
To an accuracy of better than $0.1\%$ one has
\begin{equation}\label{rhetbar}
\bar\varrho=\varrho\left(1-\frac{\lambda^2}{2}\right)\qquad
\bar\eta=\eta\left(1-\frac{\lambda^2}{2}\right)
\end{equation}
The difference between the Wolfenstein parameters 
($\varrho$, $\eta$), defined in (\ref{wodef}), and the vertex of the
normalized unitarity triangle ($\bar\varrho$, $\bar\eta$) in 
Fig. \ref{utfig} is about 2.4\%, which will have to be taken
into account in future high precision studies.
It is customary to denote the angles of the unitarity triangle
by $\alpha$, $\beta$ and $\gamma$ as shown in Fig. \ref{utfig}.

In general $\lambda$ and $A$ can be determined from decays allowed
at tree level. The parameter $\lambda$ is measured in
$K\to\pi e\nu$ or hyperon decays and $A=V_{cb}/\lambda^2$ can
be extracted from either exclusive or inclusive $b\to c$
transitions.
On the other hand, determinations of $\varrho$ and $\eta$ have
to rely largely on rare processes, which are typically loop
induced and may involve CP violation. Observables that have been
used so far to constrain these parameters, like
$\varepsilon_K$, $b\to ul\nu$ and $\Delta m_{B_d}$, suffer from
considerable theoretical uncertainties. These will ultimately limit
the accuracy of CKM determinations, even with continuing progress
on the experimental side.
In order to achieve decisive tests, it is mandatory to consider
observables where theoretical uncertainties are very well
under control.

Among the quantities best suited for this purpose are the CP
violating asymmetry in $B_d(\bar B_d)\to J/\psi K_S$,
measuring $\sin 2\beta$, and the branching ratio of 
$K_L\to\pi^0\nu\bar\nu$, determining $\eta$. These observables
have essentially no theoretical uncertainties and can be pursued at
future $B$-factories and at dedicated kaon experiments,
respectively.
\\
Some other processes such as $K^+\to\pi^+\nu\bar\nu$,
the ratio of mixing parameters $x=\Delta m/\Gamma$ in
the $B_s$ and the $B_d$ system $x_s/x_d$,
and, potentially, also CP-asymmetries in $B_d\to\pi\pi$,
have only slightly larger theoretical ambiguities. Still they are
extraordinarily clean and therefore prime candidates for
precisely testing the CKM paradigm.

The purpose of this article is to discuss the prospects for
high precision determinations of the CKM matrix using clean
observables with very small theoretical uncertainty. We consider
various strategies that will measure the CKM parameters and allow
unambiguous Standard Model tests. We compare the potential of
CP violation measurements in $B$ physics with that of 
$K_L\to\pi^0\nu\bar\nu$ and $K^+\to\pi^+\nu\bar\nu$.
Both ways allow to determine the unitarity triangle with
comparable accuracy. A combination of these complementary results
promises detailed insight into the physics of quark mixing and
CP violation.
\\
This paper is organized as follows. Theoretical uncertainties
are discussed and summarized in section 2. The subsequent section 3
reviews briefly the theoretical status of CP asymmetries in
$B_d\to J/\psi K_S$ and $B_d\to\pi\pi$ in the context of
measuring $\sin 2\beta$ and $\sin 2\alpha$. Various strategies
to determine the unitarity triangle ($\bar\varrho$, $\bar\eta$)
are considered and compared in section 4. Finally, section 5 
contains our conclusions.

\section{Theoretical Uncertainties in $K\to\pi\nu\bar\nu$}
\label{kpinunu}

The rare decays $K^+\to\pi^+\nu\bar\nu$ and
$K_L\to\pi^0\nu\bar\nu$ are loop induced FCNC processes in the
Standard Model. Being semileptonic and short-distance dominated
these channels are theoretically exceptionally well under control.
They are therefore sensitive probes of the physics at high
energy scales and allow in particular to access the CKM couplings
of the top quark in a very clean way.
In the present section we shall briefly review the theoretical
status of the $K\to\pi\nu\bar\nu$ decay modes.

\subsection{$K^+\to\pi^+\nu\bar\nu$}
\label{kplus}

The branching fraction of $K^+\to\pi^+\nu\bar\nu$ can be written
as follows  
\begin{equation}\label{bkpnn}
B(K^+\to\pi^+\nu\bar\nu)=\kappa\cdot
\left[\left(\frac{{\rm Im}\lambda_t}{\lambda^5}X(x_t)\right)^2+
\left(\frac{{\rm Re}\lambda_c}{\lambda}P_0(K^+)+
\frac{{\rm Re}\lambda_t}{\lambda^5}X(x_t)\right)^2\right]
\end{equation}
\begin{equation}\label{kapp}
\kappa=r_{K^+}\frac{3\alpha^2 B(K^+\to\pi^0 e^+\nu)}{
 2\pi^2\sin^4\Theta_W}\lambda^8=4.11\cdot 10^{-11}
\end{equation}
Here $x_t=m^2_t/M^2_W$, $\lambda_i=V^*_{is}V_{id}$ and 
$r_{K^+}=0.901$ summarizes isospin breaking corrections in relating
$K^+\to\pi^+\nu\bar\nu$ to the well measured leading decay
$K^+\to\pi^0 e^+\nu$. In the standard parametrization $\lambda_c$
is real to an accuracy of better than $10^{-3}$. The function
$X$ is given by
\begin{equation}\label{xxt}
X(x)=\eta_X\cdot \frac{x}{8}\left[\frac{x+2}{x-1}+
 \frac{3x-6}{(x-1)^2}\ln x\right]\qquad \eta_X=0.985
\end{equation} 
where $\eta_X$ is the NLO correction calculated in \cite{BB2}.
With $m_t\equiv\bar m_t(m_t)$ the QCD factor $\eta_X$ is practically
independent of $m_t$. Next
\begin{equation}\label{p0k}
P_0(K^+)=\frac{1}{\lambda^4}\left[\frac{2}{3}X^e_{NL}+
 \frac{1}{3}X^\tau_{NL}\right]
\end{equation}
represents the charm contribution
with $X^l_{NL}$ calculated in \cite{BB3}. The central value
of $P_0(K^+)$ for $\Lambda^{(4)}_{\overline{MS}}=325 MeV$,
$m_c=\bar m_c(m_c)=1.3 GeV$ and the renormalization scale $\mu_c=m_c$
is $P_0(K^+)=0.400$.
We remark that in writing $B(K^+\to\pi^+\nu\bar\nu)$ in the form
of (\ref{bkpnn}) a negligibly small term
$\sim(X^e_{NL}-X^\tau_{NL})^2$ has been omitted ($0.2\%$ effect on the
branching ratio).
\\
In general a measurement of $B(K^+\to\pi^+\nu\bar\nu)$ alone
yields a constraint on ${\rm Re}\lambda_t$ and ${\rm Im}\lambda_t$
according to eq. (\ref{bkpnn}). This relationship is very clean
and uncertainties arise only from the branching fraction, the 
charm contribution and the top quark mass, where the latter error
is almost negligible.
\\
Using in addition information from $A$ (or $V_{cb}$), the relation
between ${\rm Re}\lambda_t$ and ${\rm Im}\lambda_t$ can be
translated into a constraint in the $(\bar\varrho, \bar\eta)$
plane.
For fixed input parameters this constraint is approximately an
ellipse centered at $\bar\varrho=1+P_0(K^+)/(A^2 X(x_t))$,
$\bar\eta=0$, which is shifted from 
$(\bar\varrho, \bar\eta)=(1, 0)$ by the presence of
the charm contribution as indicated in Fig \ref{utfig}. 
\\
To learn more on the CKM parameters from
$B(K^+\to\pi^+\nu\bar\nu)$ requires additional input. One suitable
further piece of information, like $|V_{ub}/V_{cb}|$ or
$B(K_L\to\pi^0\nu\bar\nu)$, is however sufficient to determine the
CKM matrix completely. All CKM elements are then given in both
magnitude and phase, in particular $V_{td}$.

In the following we briefly address the most important uncertainties
in the theoretical treatment of $K^+\to\pi^+\nu\bar\nu$.
\begin{itemize}
\item
The top contribution is characterized by high energy scales of
${\cal O}(m_t)$ where QCD perturbation theory is a very reliable
tool. The inclusion of ${\cal O}(\alpha_s)$ corrections essentially
eliminates the sizable renormalization scale dependence of the leading 
order result. This analysis indicates that the residual uncertainty
in $X(x_t)$, for fixed $m_t$, is merely at the level of
${\cal O}(1\%)$ and thus practically irrelevant.
\item
For the charm contribution the situation is less favorable, since
QCD perturbation theory cannot be expected to be as accurate at the
rather low scale of $m_c$. This case further requires the resummation 
of large logarithms $\ln M_W/m_c$ using renormalization group
methods. Still the reliability of the calculation can be much
improved by performing a next-to-leading log analysis where,
in addition to the leading logarithms of 
${\cal O}(x_c\alpha^n_s\ln^{n+1}x_c)$, the terms of
${\cal O}(x_c\alpha^n_s\ln^{n}x_c)$ are included in the charm
quark function $X_c=X_{NL}$.
The calculation of the NLO corrections 
allows a better assessment of the applicability of perturbation theory.
In fact, the NLO correction turns out to be sufficiently small
for this approach to make sense in the present context. Furthermore,
the sensitivity to the unphysical renormalization scale
$\mu_c={\cal O}(m_c)$ is reduced at NLO. The remaining ambiguity
is to be interpreted as a theoretical uncertainty, due to the use
of a truncated perturbation series, and is about $\pm 10\%$
in $P_0(K^+)$.
\\
This ambiguity corresponds to part of the neglected higher order
corrections 
and thus provides a quantitative estimate for their order of
magnitude. Of course, knowledge of the complete
${\cal O}(x_c\alpha^{n+1}_s\ln^n x_c)$ terms, appearing at 
the order beyond next-to-leading logs, could strictly speaking give a
more rigorous estimation of the residual error. This order has
however not yet been fully calculated.
Given that the perturbative expansion for $X_{NL}$ appears rather
well behaved after renormalization group improvement and the
NLO value is within the range obtained by varying the scale
($1GeV\leq\mu_c\leq 3GeV$) in the LO result, we expect the
error estimate based on the scale dependence to give a fair account of
the actual uncertainty. In fact, additional support for this procedure
comes from considering the term ${\cal O}(x_c\alpha_s)$ in the charm
quark function $P_0(K^+)$ \cite{BB3}. This term is a contribution
beyond NLO and therefore not included in $X_{NL}$. It is
however known from the calculation of $X(x_t)$. It provides another,
independent estimate of the typical size of neglected higher order terms.
Quantitatively its size is $\sim 10\%$ in $P_0(K^+)$,
compatible with the error estimate based on the $\mu_c$-dependence.
\item
Besides through top and charm quark loops, which are short-distance
in character due to $m_t$, $m_c\gg\Lambda_{QCD}$,
$K^+\to\pi^+\nu\bar\nu$ may also proceed through second order 
weak interactions involving up quarks. This mechanism is the source
of long-distance contributions to $K\to\pi\nu\bar\nu$, which are 
determined by nonperturbative low-energy QCD dynamics and
difficult to calculate reliably. Of crucial importance for the
high accuracy that can be achieved in the theoretical treatment of
$K\to\pi\nu\bar\nu$ is the fact that such contributions are very
small. The reason for this is a hard GIM suppression of the
electroweak $\bar sd\to\nu\bar\nu$ amplitude. This means that
the charm contribution behaves as $m^2_c\ln M_W/m_c$ for $m_c\to 0$,
rather than, say, just logarithmically $\sim\ln M_W/m_c$. Hence the
size of the short-distance dominated charm sector is essentially
determined by $m^2_c$, while the long distance up quark contribution
is characterized by the QCD scale $\Lambda^2_{QCD}$ (the up
quark mass being negligible). The long distance part is therefore
suppressed by $\Lambda^2_{QCD}/m^2_c$ relative to the charm quark
amplitude. Detailed estimates \cite{RS,HL,LW,FAJ} quantify this general
suppression pattern. Using chiral perturbation theory
the authors of \cite{LW} find 
$X_{LD}\;\raisebox{-.4ex}{\rlap{$\sim$}} \raisebox{.4ex}{$<$}\;
(g_8\pi^2/3)(f_\pi/M_W)^2$, with $g_8=5.1$.
This estimate is based on the amplitude involving one $W$-
and one $Z$-boson exchange, which is enhanced by the
$\Delta I=1/2$ rule and can be expected to be dominant. The result
is less than about $5\%$ of the charm contribution and negligible
in view of the perturbative uncertainty in the charm sector.
A further long distance mechanism, involving two $W$-boson
exchanges, is $K^+\to\nu_l l^{+*}\to\nu_l\pi^+\bar\nu_l$,
$l=e,\mu$, calculated in \cite{HL}. It amounts to
$\approx 2\%$ of the charm amplitude and is likewise negligible.
Similar conclusions on the long distance contribution have been
reached in \cite{RS,FAJ}.
\item
To eliminate the hadronic matrix element 
$\langle\pi|(\bar sd)_V|K\rangle$ in the calculation of
$K^+\to\pi^+\nu\bar\nu$, the branching ratio
$B(K^+\to\pi^+\nu\bar\nu)$ can be related to
$B(K^+\to\pi^0 e^+\nu)$ using isospin symmetry.
Corrections to the strict isospin limit have been considered
in \cite{MP}. They arise from phase space effects due to differences
in the mass of $\pi^+$ and $\pi^0$ (or $K^0$ and $K^+$ for the
neutral mode $K_L\to\pi^0\nu\bar\nu$), isospin violation in the
$K\to\pi$ form factors, and electromagnetic radiative corrections
that affect the $\bar s\to\bar u e^+\nu$ transition, but not
$\bar s\to\bar d\nu\bar\nu$. Ultimately these effects stem from the
usual sources of isospin breaking, the electromagnetic interaction
or the $u$-$d$ mass difference.
For the correction factor $r_{K^+}$ in (\ref{kapp}),
\cite{MP} obtain $r_{K^+}=0.9614\cdot 0.9574\cdot 0.979=0.901$,
where the first factor is from phase space, the second from the
$K\to\pi$ form factors and the last from QED radiative
corrections. Since the meson masses are known precisely the phase
space effect has essentially no uncertainty.
The QED correction factor is calculated in the leading log
approximation and given by
$(1+2\alpha/\pi \ln(M_Z/\mu_h))^{-1}=0.979$ for 
$\alpha=1/137$ and $\mu_h=m_p=0.938GeV$. Taking into account
the various ambiguities in this calculation, from
non-logarithmic ${\cal O}(\alpha)$ corrections, the fact that 
$\alpha$ could be $\alpha(M_Z)$ rather than $\alpha(m_e)$,
replacing $M_Z$ by $M_W$ or varying $\mu_h$ between
$0.5GeV$ and $2GeV$, one finds typically an uncertainty of
$\pm 0.5\%$.
Finally, the expression for $B(K^+\to\pi^+\nu\bar\nu)$ receives
a small error from the use of $B(K^+\to\pi^0 e^+\nu)=0.0482$,
which is currently measured to $1\%$ accuracy.
\end{itemize}

To summarize, the theoretical uncertainty in 
$K^+\to\pi^+\nu\bar\nu$ is dominated by the charm contribution.
The latter is estimated to be $P_0(K^+)=0.40\pm 0.047$,
where the error bar represents the symmetrized range obtained by
varying the renormalization scale $\mu_c$ between $1 GeV$ and
$3 GeV$. This uncertainty translates into a $\pm 5\%$ variation
of the branching ratio. The other errors, such as those from the 
scale dependence in the top sector or from the long distance 
contribution, are small in comparison and can be neglected.
\\
The intrinsic theoretical uncertainties we have discussed so
far should be distinguished from uncertainties in basic
Standard Model parameters.
Among these are the errors in $V_{cb}$ and $m_t$ that will
be specified later on. In the charm sector one has the charm quark mass 
and the QCD scale for which we shall take 
$m_c=\bar m_c(m_c)=(1.30\pm 0.05)GeV$ 
(the running ${\overline{MS}}$ mass) and
$\Lambda^{(4)}_{\overline{MS}}=(325\pm 75) MeV$.
Here we have anticipated that by the time $K^+\to\pi^+\nu\bar\nu$
will be measured, the precision in $m_c$ should have improved
over the curent status. Combining the uncertainties from theory,
$m_c$ and $\Lambda$ we finally obtain
\begin{equation}\label{pknum}
P_0(K^+)=0.400\pm 0.047\ (th)\ \pm 0.035\ (m_c)\ \pm 0.026\
(\Lambda)\ =0.40\pm 0.06
\end{equation}
which we will use in the analysis below.

\subsection{$K_L\to\pi^0\nu\bar\nu$}
\label{klong}

Due to the CP properties of $K_L$, $\pi^0$ and the relevant
hadronic, short-distance transition current, the mode
$K_L\to\pi^0\nu\bar\nu$ proceeds in the SM almost entirely
through direct CP violation.
In explicit terms the branching fraction per neutrino flavor is given
by
\begin{equation}\label{bklpnl}
B(K_L\to\pi^0\nu_l\bar\nu_l)=r_{K_L}\frac{\tau_{K_L}}{\tau_{K^+}}
\frac{\alpha^2 B(K^+\to\pi^0 e^+\nu)}{2\pi^2\sin^4\Theta_W|V_{us}|^2}
\cdot\frac{1}{2}
\frac{\left|\xi-\xi^*\frac{1-\bar\varepsilon}{1+\bar\varepsilon}\right|^2}{
1+\left|\frac{1-\bar\varepsilon}{1+\bar\varepsilon}\right|^2}
\end{equation}
Here $\xi=\sum_{i=u,c,t}\lambda_i X_i$ and $r_{K_L}=0.944$ is the
isospin breaking correction \cite{MP} from relating 
$K_L\to\pi^0\nu\bar\nu$ to $K^+\to\pi^0 e^+\nu$. The factor
\begin{equation}\label{epsbar}
\frac{1-\bar\varepsilon}{1+\bar\varepsilon}=
\frac{M^*_{12}-i\Gamma^*_{12}/2}{(\Delta m-i\Delta\Gamma/2)/2}
\end{equation}
derives from $|K_L\rangle\sim (1+\bar\varepsilon)|K^0\rangle+
(1-\bar\varepsilon)|\bar K^0\rangle$, with $M_{12}$ and
$\Gamma_{12}$ denoting the off-diagonal elements in the neutral
kaon mass- and decay constant matrix, respectively.
$\Delta m=m_L-m_S$ ($\Delta\Gamma=\Gamma_L-\Gamma_S$) is the
difference in mass (decay rate) between the eigenstates $K_L$
and $K_S$. We use the CP phase conventions
$CP|K^0\rangle=-|\bar K^0\rangle$,
$CP(\bar ds)_V CP^{-1}=-(\bar sd)_V$. (The neutral pion has
negative CP parity $CP|\pi^0\rangle=-|\pi^0\rangle$.)
\\
In principle arbitrary phases could be introduced in the CP
transformation of $K^0$ and the current $(\bar ds)_V$. These phases
would multiply the factor $\xi^*$ in (\ref{bklpnl}).
However, compensating phases would then be present in the hadronic 
matrix elements of $M_{12}$, $\Gamma_{12}$, assuring that the
physics remains unchanged.
Note further that the expression in (\ref{bklpnl}) is manifestly
invariant under rephasing of the quark fields, since
$(1-\bar\varepsilon)/(1+\bar\varepsilon)\sim\lambda^2_i$.
In particular one has
$(\lambda^*_u/\lambda_u)(1-\bar\varepsilon)/(1+\bar\varepsilon)=1$
up to a few times $10^{-3}$, which is independent of the CKM matrix
phase convention. It then follows that
\begin{equation}\label{imltlus}
\left|\xi-\xi^*\frac{1-\bar\varepsilon}{1+\bar\varepsilon}\right|^2=
\left|\xi-\xi^*\frac{\lambda_u}{\lambda^*_u}\right|^2=
4\frac{({\rm Im}\lambda_t\lambda^*_u)^2}{|\lambda_u|^2}
(X_t-X_c)^2
\end{equation}
where we have neglected long distance contributions.
This expression is manifestly rephasing invariant.
\\
Neglecting the charm quark contribution, which affects the branching
ratio by only $0.1\%$, specializing to the standard CKM parametrization
where $\lambda^*_u=\lambda_u$, and summing over the three neutrino
species, one obtains the familiar result
\begin{equation}\label{bklpn}
B(K_L\to\pi^0\nu\bar\nu)=\kappa_L\cdot
\left(\frac{{\rm Im}\lambda_t}{\lambda^5}X(x_t)\right)^2
\qquad {\rm Im}\lambda_t=\eta A^2\lambda^5
\end{equation}
\begin{equation}\label{kapl}
\kappa_L=r_{K_L}\frac{\tau_{K_L}}{\tau_{K^+}}
\frac{3\alpha^2 B(K^+\to\pi^0 e^+\nu)}{2\pi^2\sin^4\Theta_W}
\lambda^8=1.80\cdot 10^{-10}
\end{equation}
Equation (\ref{bklpn}) provides a very accurate relationship between
the observable $B(K_L\to\pi^0\nu\bar\nu)$ and fundamental SM
parameters. The high precision that can be achieved in the theoretical
calculation of this decay mode is rather unique among rare decay
phenomena.

\begin{itemize}
\item
$K_L\to\pi^0\nu\bar\nu$ shares many features with the charged
mode $K^+\to\pi^+\nu\bar\nu$, which make it already a very clean
process. This situation is still improved considerably by the
CP violating nature of $K_L\to\pi^0\nu\bar\nu$, since here only
the top contribution is significant and all the uncertainties
associated with the charm sector are eliminated. After including
NLO corrections, the intrinsic theoretical uncertainty in
$X^2(x_t)$ from truncating the perturbation series is estimated
to be $\pm 1\%$.
\item
Long distance contributions to $K_L\to\pi^0\nu\bar\nu$ are still
further suppressed compared to the case of $K^+\to\pi^+\nu\bar\nu$
due to the CP violating property of the neutral mode. They are
likewise completely negligible \cite{RS}.
\item
The factor $B(K^+\to\pi^0 e^+\nu)\tau_{K_L}/\tau_{K^+}$ serves
to eliminate the hadronic matrix element required for the
calculation of $K_L\to\pi^0\nu\bar\nu$. The combined experimental
error in this quantity is $\pm 1.5\%$, dominated by the
uncertainty in $B(K^+\to\pi^0 e^+\nu)$ \cite{PDG}. This error
can be further reduced by improved measurements in the future.
\item
The isospin breaking correction $r_{K_L}$ is here
$r_{K_L}=1.0522\cdot 0.9166\cdot 0.979=0.944$ \cite{MP}.
The first factor comes from the difference in phase space between
$K^0\to\pi^0$ and $K^+\to\pi^0$ decay and does not introduce any 
significant error. The short-distance QED correction 0.979 is the same as 
in the case of $K^+\to\pi^+\nu\bar\nu$ and has an uncertainty of
probably below $\pm 0.5\%$.
\item
{}From the full expression given in (\ref{bklpnl}) one can derive the
contribution of indirect CP violation to $B(K_L\to\pi^0\nu\bar\nu)$.
For this purpose it is convenient to use the CKM phase conventions of 
the standard parametrization. We further approximate
\begin{equation}\label{epseps}
\bar\varepsilon\approx\varepsilon=\frac{1+i}{\sqrt{2}}|\varepsilon|
\qquad |\varepsilon|=(2.282\pm 0.019)\cdot 10^{-3}
\end{equation}
where $\varepsilon$ is the parameter describing indirect CP
violation in $K^0\to\pi\pi$ decays \cite{PDG}.
Expanding to first order in $|\varepsilon|$ one finds that the effect
of indirect CP violation in $K_L\to\pi^0\nu\bar\nu$ is to multiply
the branching ratio in (\ref{bklpn}) by a factor of
\begin{equation}\label{cpindir}
1+\sqrt{2}|\varepsilon|\frac{{\rm Re}\xi}{{\rm Im}\xi}\quad
{\rm where}\quad
\frac{{\rm Re}\xi}{{\rm Im}\xi}=-
 \frac{1+\frac{P_0(K^+)}{A^2X(x_t)}-\varrho}{\eta}
\end{equation}
Since ${\rm Re}\xi/{\rm Im}\xi$ is typically $-4$, we find that
indirect CP violation reduces the branching fraction
$B(K_L\to\pi^0\nu\bar\nu)$ by $\approx 1\%$. We shall neglect this
small correction for simplicity. The effect can of course be taken
into account in the future, should such a high precision be required.
\end{itemize}

\section{CP Asymmetries in $B_d$ Decays}
\label{CPB}

The observation of CP violating asymmetries in neutral $B$ decays
to CP eigenstates will test the Standard Model and allow to determine
angles of the unitarity triangle in Fig. \ref{utfig}. Among the
most promising candidates for these experiments are the decay mode
$B_d(\bar B_d)\to J/\psi K_S$ and, to a lesser extent, also
$B_d(\bar B_d)\to\pi^+\pi^-$, which will be pursued in particular
at the upcoming $B$-factories. The corresponding time dependent
or time integrated (at hadron colliders)
CP asymmetries in the decay of tagged $B_d$ compared to $\bar B_d$,
measure $\sin 2\beta$ and $\sin 2\alpha$, respectively.
This subject has been extensively discussed in the literature.
Here we content ourselves with recalling a particular aspect, the 
effect of penguin contributions, which is important for the theoretical
accuracy in infering $\sin 2\phi$, $\phi=\alpha$, $\beta$, from
measured asymmetries. In the absence of a penguin amplitude the
time dependent asymmetry oscillates as $\sin\Delta m_{B_d}t$ with an 
amplitude given by $\sin 2\phi$. When a small penguin contribution is
present in addition to the dominant tree level amplitude, the amplitude
of $\sin\Delta m t$ does not in general measure $\sin 2\phi$
alone, but the combination \cite{GRO1}
\begin{equation}\label{sin2p}
\sin 2\phi- 2\left|\frac{A_2}{A_1}\right|\cos 2\phi
\cos(\delta_1-\delta_2) \sin(\phi_1-\phi_2)
\end{equation}
where $A_i$, $\delta_i$ and $\phi_i$ are the amplitude, the strong
phase and the weak phase, respectively, of the tree ($i=1$) and
the penguin contribution ($i=2$) for $B_d\to f$. The strong phases
are unknown and $\cos(\delta_1-\delta_2)$ could be one in the
worst case.
\\
The ratio of $|A_2/A_1|$ is expected to be typically 
${\cal O}(3-5\%)$ for $B_d\to J/\psi K_S$ and
${\cal O}(10-20\%)$ for $B_d\to\pi^+\pi^-$. The penguin amplitude
is slightly enhanced in the latter case through the ratio of
CKM angles $|V^*_{tb}V_{td}/(V^*_{ub}V_{ud})|\sim 3$, whereas
in $B_d\to J/\psi K_S$ this factor is
$|V^*_{tb}V_{ts}/(V^*_{cb}V_{cs})|\sim 1$.
It should be remarked that these estimates of $|A_2/A_1|$ are highly
uncertain due to the poor knowledge of hadronic matrix elements.
\\
For $B_d\to J/\psi K_S$ this potential problem is however practically
eliminated since the tree- and the penguin amplitude have almost
identical weak phases. More quantitatively, 
$\sin(\phi_1-\phi_2)\simeq\lambda^2\eta\approx 0.02$
and the penguin contamination in (\ref{sin2p}) is estimated to be
below $\pm 0.002$.
\\
The situation is not as fortunate for $B_d\to\pi^+\pi^-$, where
$\sin(\phi_1-\phi_2)=\sin\alpha$. As pointed out in \cite{GRO2},
if $\alpha\approx\pi/2$, which can not be excluded at present,
$\sin 2\alpha\approx 0$. However, the asymmetry coefficient
(\ref{sin2p}), which is supposed to measure $\sin 2\alpha$,
could at the same time be as large as $\sim 0.4$ due to penguin
effects. For larger $\sin 2\alpha$ the impact of the penguin
contribution is smaller. A detailed discussion can be found in
\cite{GRO2}. More recently, this problem has also been addressed
in \cite{KPW}.

As shown in \cite{GL,NQ} the penguin contamination could be eliminated
in principle by an isospin analysis. This however requires the 
measurement of the rates for $B^+\to\pi^+\pi^0$ and
$B_d\to\pi^0\pi^0$, and their CP conjugates, 
which will be difficult to achieve, in particular in view of the
fact that the branching ratio for $B_d\to\pi^0\pi^0$ is
expected to be below $10^{-6}$ \cite{KP}.
\\
In summary, while CP asymmetries in $B_d\to J/\psi K_S$ are a very
clean measure of $\sin 2\beta$, the extraction of $\sin 2\alpha$
{}from $B_d\to\pi^+\pi^-$ is somewhat more problematic. If the 
difficulties related to penguin contributions can be overcome, also
this channel will be a very useful observable for CKM matrix
determinations. A recent discussion of alternative methods for
the extraction of $\alpha$ can be found in \cite{RF}.

\section{Determinations of the Unitarity Triangle}
\label{utdet}

We shall now describe several applications of the observables we
have discussed above for precise determinations of the CKM matrix.
Four independent pieces of information are needed to fix the four
parameters of quark mixing $\lambda$, $A$, $\varrho$ and $\eta$.
This determination is the necessary first step towards a comprehensive
test of this important Standard Model sector. Those
physical quantities should be chosen for this purpose, that allow 
to define the most accurate set of CKM parameters and therefore
constitute a firm basis for any further tests and comparisons. Which
observables will eventually turn out to provide the optimal set
of CKM matrix input is not yet completely clear at present, but
theoretically clean processes like $K\to\pi\nu\bar\nu$ and the
CP asymmetries, both time dependent and time integrated, in
the "gold-plated" mode $B_d\to J/\psi K_S$ are certainly prime 
candidates.
\\
In the following we illustrate several scenarios for determining
the CKM matrix and show what degree of accuracy can be expected 
in the future.

\subsection{Unitarity Triangle from $K\to\pi\nu\bar\nu$ and
from $\sin 2\alpha$ and $\sin 2\beta$}
\label{utks2ab}

The most obvious source for two of the parameters are weak
decays allowed at tree level: $K\to\pi e\nu$ and hyperon decays
give $\lambda$ and $A=V_{cb}/\lambda^2$ can be extracted from
exclusive and inclusive semileptonic $b\to c$ transitions.
Measuring $\sin 2\alpha$ and $\sin 2\beta$ from CP asymmetries in
$B$ decays allows, in principle, to fix the remaining two
parameters $\bar\eta$ and $\bar\varrho$, which can be expressed as
\cite{B94}
\begin{equation}\label{ersab}
\bar\eta=\frac{r_-(\sin 2\alpha)+r_+(\sin 2\beta)}{1+
  r^2_+(\sin 2\beta)}\qquad
\bar\varrho=1-\bar\eta r_+(\sin 2\beta)
\end{equation}
where $r_\pm(z)=(1\pm\sqrt{1-z^2})/z$.
In general the calculation of $\bar\varrho$ and $\bar\eta$ from
$\sin 2\alpha$ and $\sin 2\beta$ involves discrete ambiguities.
As described in \cite{B94}
they can be resolved by using further information, e.g. bounds on
$|V_{ub}/V_{cb}|$, so that eventually the solution (\ref{ersab})
is singled out.
\\
Alternatively, $\bar\varrho$ and $\bar\eta$ may also be determined
{}from $K^+\to\pi^+\nu\bar\nu$ and $K_L\to\pi^0\nu\bar\nu$ alone
\cite{BH,BB4}. An interesting feature of this possibility is in
particular that the extraction of $\sin 2\beta$ from these
two modes is essentially independent of $m_t$ and $V_{cb}$
\cite{BB4}. This fact enables a rather accurate determination of
$\sin 2\beta$ from $K\to\pi\nu\bar\nu$.

A comparison of both strategies
is displayed in Table \ref{tabkb}, where 
the following input has been used
\begin{equation}\label{vcbmt}
V_{cb}=0.040\pm 0.002\qquad m_t=(170\pm 3) GeV
\end{equation}
\begin{equation}\label{bklkp}
B(K_L\to\pi^0\nu\bar\nu)=(3.0\pm 0.3)\cdot 10^{-11}\qquad
B(K^+\to\pi^+\nu\bar\nu)=(1.0\pm 0.1)\cdot 10^{-10}
\end{equation}
The charm contribution in $K^+\to\pi^+\nu\bar\nu$ is assumed to
be known to $\pm 15\%$, $P_0(K^+)=0.40\pm 0.06$.
\\
The measurements of CP asymmetries in $B_d\to\pi\pi$ and
$B_d\to J/\psi K_S$, expressed in terms of $\sin 2\alpha$ and
$\sin 2\beta$, are taken to be
\begin{equation}\label{sin2a2bI}
\sin 2\alpha=0.40\pm 0.10 \qquad \sin 2\beta=0.70\pm 0.06
\qquad ({\rm scenario\ I})
\end{equation}
\begin{equation}\label{sin2a2bII}
\sin 2\alpha=0.40\pm 0.04 \qquad \sin 2\beta=0.70\pm 0.02
\qquad ({\rm scenario\ II})
\end{equation}
Scenario I corresponds to the accuracy being aimed for at $B$-factories
prior to the LHC era. An improved precision can be anticipated from
LHC experiments, which we illustrate with our choice of scenario II.

As can be seen in Table \ref{tabkb}, the CKM determination
using $K\to\pi\nu\bar\nu$ is competitive with the one based
on CP violation in $B$ decays, except for $\bar\varrho$ which
is less constrained by the rare kaon processes.
\begin{table}
\begin{center}
\begin{tabular}{|c||c|c|c|}\hline
&$K\to\pi\nu\bar\nu$&$B\to\pi\pi, J/\psi K_S$ (I) 
&$B\to\pi\pi, J/\psi K_S$ (II) \\
\hline
\hline
$|V_{td}|/10^{-3}$&$10.3\pm 1.1(\pm 0.9)$&$8.8\pm 0.5(\pm 0.3)$ 
&$8.8\pm 0.5(\pm 0.2)$ \\
\hline
$|V_{ub}/V_{cb}|$&$0.089\pm 0.017(\pm 0.011)$
&$0.087\pm 0.009(\pm 0.009)$&$0.087\pm 0.003(\pm 0.003)$ \\
\hline 
$\bar\varrho$&$-0.10\pm 0.16(\pm 0.12)$&$0.07\pm 0.03(\pm 0.03)$
&$0.07\pm 0.01(\pm 0.01)$ \\
\hline
$\bar\eta$&$0.38\pm 0.04(\pm 0.03)$&$0.38\pm 0.04(\pm 0.04)$
&$0.38\pm 0.01(\pm 0.01)$ \\
\hline
$\sin 2\beta$&$0.62\pm 0.05(\pm 0.05)$&$0.70\pm 0.06(\pm 0.06)$
&$0.70\pm 0.02(\pm 0.02)$ \\
\hline
${\rm Im}\lambda_t/10^{-4}$&$1.37\pm 0.07(\pm 0.07)$
&$1.37\pm 0.19(\pm 0.15)$&$1.37\pm 0.14(\pm 0.08)$ \\
\hline
\end{tabular}
\end{center}
\caption[]{Illustrative example of the determination of CKM
parameters from $K\to\pi\nu\bar\nu$ and from CP violating
asymmetries in $B$ decays. The relevant input is as described
in the text. Shown in brackets are the errors one obtains
using $V_{cb}=0.040\pm 0.001$ instead of $V_{cb}=0.040\pm 0.002$.
\label{tabkb}}
\end{table}
On the other hand ${\rm Im}\lambda_t$ is better determined
in the kaon scenario. It can be obtained from
$K_L\to\pi^0\nu\bar\nu$ alone and does not require knowledge
of $V_{cb}$ which enters ${\rm Im}\lambda_t$ when derived
{}from $\sin 2\alpha$ and $\sin 2\beta$.
We have displayed the extraction of ${\rm Im}\lambda_t$ from
CP asymmetries in $B$ decays in more detail in Table \ref{tabimltb}.

This should be compared with the results 
for ${\rm Im}\lambda_t$ that could be obtained using
$B(K_L\to\pi^0\nu\bar\nu)$. Taking
$B(K_L\to\pi^0\nu\bar\nu)=(3.0\pm 0.3)\cdot 10^{-11}$ and
$m_t=(170\pm 3)GeV$ (case ($a$)), and 
$B(K_L\to\pi^0\nu\bar\nu)=(3.0\pm 0.15)\cdot 10^{-11}$ and
$m_t=(170\pm 1)GeV$ (case ($b$)), we find
\begin{equation}\label{imlta}
{\rm Im}\lambda_t/10^{-4}=1.368\pm 0.069\pm 0.028=
1.368\pm 0.074\quad (a)
\end{equation}
\begin{equation}\label{imltb}
{\rm Im}\lambda_t/10^{-4}=1.368\pm 0.035\pm 0.009=
1.368\pm 0.036\quad (b)
\end{equation}
\begin{table}
\begin{center}
\begin{tabular}{|c|c||c|c|c||c|}\hline
&&$\Delta(\sin 2\alpha)$&$\Delta(\sin 2\beta)$&
  $\Delta(V_{cb})$&$\Delta_{total}$ \\
\hline
\hline
I&1.370&$\pm 0.030$&$\pm 0.131$&$\pm 0.137$ [$\pm 0.069$]&
 $\pm 0.192$ [$\pm 0.151$] \\
\hline
II&1.370&$\pm 0.012$&$\pm 0.044$&$\pm 0.137$ [$\pm 0.069$]&
 $\pm 0.144$ [$\pm 0.083$] \\
\hline
\end{tabular}
\end{center}
\caption[]{${\rm Im}\lambda_t/10^{-4}$ as determined from CP
asymmetries in $B$ decays. Scenario I assumes
$\sin 2\alpha=0.40\pm 0.10$, $\sin 2\beta=0.70\pm 0.06$.
For scenario II we take
$\sin 2\alpha=0.40\pm 0.04$, $\sin 2\beta=0.70\pm 0.02$.
We use $V_{cb}=0.040\pm 0.002$ and, for the results in square brackets,
$V_{cb}=0.040\pm 0.001$.
\label{tabimltb}}
\end{table}
The comparison suggests that $K_L\to\pi^0\nu\bar\nu$ should eventually 
yield the most accurate value of ${\rm Im}\lambda_t$.
This would be an important result since ${\rm Im}\lambda_t$
plays a central role in the phenomenology of CP violation
in $K$ decays and is furthermore equivalent to the 
Jarlskog parameter $J_{CP}$ \cite{CJ}, 
the invariant measure of CP violation in the Standard Model, 
$J_{CP}=\lambda(1-\lambda^2/2){\rm Im}\lambda_t$.

\subsection{Unitarity Triangle from $K_L\to\pi^0\nu\bar\nu$
and $\sin 2\alpha$}

\begin{figure}[ht]
   \vspace{-4cm}
   \epsfxsize=10cm
   \centerline{\epsffile{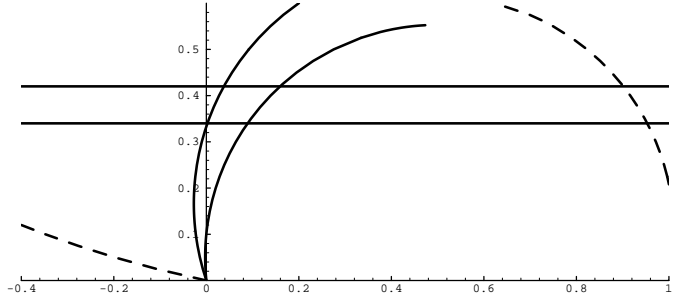}}
   \vspace{-4cm}
\caption{\label{s2aklfig} Constraints in the $(\bar\varrho, \bar\eta)$
 plane from $\sin 2\alpha=0.4\pm 0.2$ ("half-moon") and from
 $B(K_L\to\pi^0\nu\bar\nu)$ (horizontal band, input as
 specified in eqns. (18) and (19)).
 The dashed curves 
 illustrate the discrete ambiguities involved in determining
 $\bar\varrho$ and $\bar\eta$ from $\sin 2\alpha$
 for the central value $\sin 2\alpha=0.4$. They can be eliminated
 by information on $\varepsilon_K$ and $|V_{ub}/V_{cb}|$.
 Note that even for a quite loosely determined $\sin 2\alpha$ as in 
 the present example, the resulting constraint in the 
 $(\bar\varrho, \bar\eta)$ plane is rather tight.}
\end{figure}

Next, results from CP asymmetries in $B$ decays could also be
combined with measurements of $K\to\pi\nu\bar\nu$.
As an illustration we would like to discuss a scenario where
the unitarity triangle is determined by $\lambda$, $V_{cb}$,
$\sin 2\alpha$ and $B(K_L\to\pi^0\nu\bar\nu)$
(see Fig. \ref{s2aklfig}).
In this case $\bar\eta$ follows directly from 
$B(K_L\to\pi^0\nu\bar\nu)$ (\ref{bklpn}) and $\bar\varrho$ is
obtained using \cite{B94}
\begin{equation}\label{rhoalpha}
\bar\varrho=\frac{1}{2}-\sqrt{\frac{1}{4}-\bar\eta^2+
\bar\eta r_-(\sin 2\alpha)}
\end{equation}
where $r_-(z)$ is defined after eq. (\ref{ersab}).
The advantage of this strategy is that most CKM quantities are
not very sensitive to the precise value of $\sin 2\alpha$.
Moreover a high accuracy in the Jarlskog parameter and in
${\rm Im}\lambda_t$ is automatically guaranteed. As shown in
Table \ref{tabkl2a}, very respectable results can be expected
for other quantities as well, with only modest requirements
on the accuracy of $\sin 2\alpha$. 
It is conceivable that theoretical uncertainties due to penguin
contributions could eventually be brought under control at least
to the level assumed in Table \ref{tabkl2a}. 
\begin{table}
\begin{center}
\begin{tabular}{|c||c|c|c|}\hline
&&A&B \\
\hline
\hline
$\bar\eta$&$0.380$&$\pm 0.043$&$\pm 0.028$ \\
\hline
$\bar\varrho$&$0.070$&$\pm 0.058$&$\pm 0.031$ \\
\hline
$\sin 2\beta$&$0.700$&$\pm 0.077$&$\pm 0.049$ \\
\hline
$|V_{td}|/10^{-3}$&$8.84$&$\pm 0.67$&$\pm 0.34$ \\
\hline
$|V_{ub}/V_{cb}|$&$0.087$&$\pm 0.012$&$\pm 0.007$ \\
\hline 
\end{tabular}
\end{center}
\caption[]{Determination of the CKM matrix from $\lambda$, $V_{cb}$,
$K_L\to\pi^0\nu\bar\nu$ and $\sin 2\alpha$ from the CP asymmetry
in $B_d\to\pi^+\pi^-$. Scenario A (B) assumes
$V_{cb}=0.040\pm 0.002 (\pm 0.001)$
and $\sin 2\alpha=0.4\pm 0.2 (\pm 0.1)$. In both cases we take
$B(K_L\to\pi^0\nu\bar\nu)\cdot 10^{11}=3.0\pm 0.3$ and
$m_t/GeV=170\pm 3$. 
\label{tabkl2a}}
\end{table}
As an alternative, $\sin 2\beta$ from $B_d\to J/\psi K_S$ 
could be used as independent input instead of $\sin 2\alpha$.
Unfortunately the combination of $K_L\to\pi^0\nu\bar\nu$ and
$\sin 2\beta$ tends to yield somewhat less restrictive constraints
on the unitarity triangle. On the other hand it has of course the
advantage of being practically free of any theoretical uncertainties.

\subsection{Unitarity Triangle and $V_{cb}$ from $\sin 2\alpha$,
$\sin 2\beta$ and $K_L\to\pi^0\nu\bar\nu$}

In \cite{B94} an additional strategy has been proposed that
could offer unprecedented precision for all basic CKM
parameters. While $\lambda$ is obtained as usual from
$K\to\pi e\nu$, $\bar\varrho$ and $\bar\eta$ could be determined
{}from $\sin 2\alpha$ and $\sin 2\beta$ as measured in CP
violating asymmetries in $B$ decays. Given $\eta$, one could
take advantage of the very clean nature of $K_L\to\pi^0\nu\bar\nu$
to extract $A$ or, equivalently $V_{cb}$. This determination
benefits further from the very weak dependence of $A$ on
the $K_L\to\pi^0\nu\bar\nu$ branching ratio, which is only with
a power of $0.25$. Moderate accuracy in $B(K_L\to\pi^0\nu\bar\nu)$
would thus still give a high precision in $V_{cb}$.
As an example we take $\sin 2\alpha=0.40\pm 0.04$,
$\sin 2\beta=0.70\pm 0.02$ and 
$B(K_L\to\pi^0\nu\bar\nu)=(3.0\pm 0.3)\cdot 10^{-11}$,
$m_t=(170\pm 3)GeV$. 
This yields
\begin{equation}\label{rhetvcb}
\bar\varrho=0.07\pm 0.01\qquad
\bar\eta=0.38\pm 0.01\qquad
V_{cb}=0.0400\pm 0.0013
\end{equation}
which would be a truly remarkable result.

\subsection{$B\to X_{d,s}\nu\bar\nu$, $B_{d,s}\to\mu^+\mu^-$
and $x_d/x_s$}

Finally we would like to mention a few additional observables,
that are theoretically very well under control and which are
therefore also potential candidates for precise CKM determinations.
These are the ratios
\begin{equation}\label{bxnn}
\frac{B(B\to X_d\nu\bar\nu)}{B(B\to X_s\nu\bar\nu)}=
\left|\frac{V_{td}}{V_{ts}}\right|^2
\end{equation}
\begin{equation}\label{bmumu}
\frac{B(B_d\to\mu^+\mu^-)}{B(B_s\to\mu^+\mu^-)}=
\frac{\tau_{B_d}}{\tau_{B_s}}\frac{m_{B_d}}{m_{B_s}}
\frac{f^2_{B_d}}{f^2_{B_s}}
\left|\frac{V_{td}}{V_{ts}}\right|^2
\end{equation}
\begin{equation}\label{xds}
\frac{x_d}{x_s}=
\frac{\tau_{B_d}}{\tau_{B_s}}\frac{m_{B_d}}{m_{B_s}}
\frac{B_{B_d}f^2_{B_d}}{B_{B_s}f^2_{B_s}}
\left|\frac{V_{td}}{V_{ts}}\right|^2
\end{equation}
which all measure 
\begin{equation}\label{vtdts}
\left|\frac{V_{td}}{V_{ts}}\right|^2=\lambda^2
\frac{(1-\bar\varrho)^2+\bar\eta^2}{1+\lambda^2(2\bar\varrho-1)}
\end{equation}
The cleanest quantity is (\ref{bxnn}), which is essentially free
of hadronic uncertainties. Next comes (\ref{bmumu}), involving
$SU(3)$ breaking effects in the ratio of $B$ meson decay constants.
Finally, $SU(3)$ breaking in the ratio of bag parameters
$B_{B_d}/B_{B_s}$ enters in addition in (\ref{xds}). These 
$SU(3)$ breaking effects should eventually be calculable with
reasonable precision from lattice QCD.
\\
In order to extract $|V_{td}|$ from either of the quantities in
(\ref{bxnn}) -- (\ref{xds}) with an accuracy competitive to the
one in the first column of Table \ref{tabkb}, the combined
theoretical and experimental uncertainty for
$|V_{td}/V_{ts}|^2$, as determined from (\ref{bxnn}) -- (\ref{xds}),
should be brought below $\pm 20\%$.

\section{Conclusions}
\label{concl}

We have discussed the phenomenological potential of theoretically
clean observables that promise to provide precise determinations
of the CKM matrix and detailed tests of Standard Model
flavordynamics. CP violation experiments at $e^+e^-$ $B$-factories
and hadron colliders will pursue the measurement of
$\sin 2\beta$ and $\sin 2\alpha$. The former is essentially free
of theoretical uncertainties and the latter will also give important
and rather clean information, provided the penguin contributions
can be sufficiently well controlled.
\\
Besides this class of phenomena the following, theoretically
very clean processes can give additional pieces of information
that will be crucial for concise tests of the CKM description
of quark mixing:
\begin{enumerate}
\begin{enumerate}
\item $B(K_L\to\pi^0\nu\bar\nu)$
\item $B(B\to X_d\nu\bar\nu)/B(B\to X_s\nu\bar\nu)$
\item $B(K^+\to\pi^+\nu\bar\nu)$
\item $B(B_d\to\mu^+\mu^-)/B(B_s\to\mu^+\mu^-)$
\item $x_d/x_s$
\end{enumerate}
\end{enumerate}
This list is essentially ordered according to increasing
theoretical uncertainties. In principle quantity (b) has basically,
like $B(K_L\to\pi^0\nu\bar\nu)$, no such uncertainties, but is
presumably even more difficult to measure.
\\
We have considered several strategies to determine the CKM
matrix. In particular we have pointed out that a measurement
of $\sin 2\alpha$ with only rather moderate precision combined with
$B(K_L\to\pi^0\nu\bar\nu)$ could give a very respectable
determination of CKM parameters. This underlines the great
importance to also succeed in measuring $\sin 2\alpha$.
\\
Since the number of theoretically clean processes is quite limited,
it is mandatory that all of them are being pursued experimentally
as far as possible, irrespective of which quantities will
ultimately turn out to give the best determination of CKM parameters.
After all the goal is not just to measure but eventually to
overconstrain the CKM matrix. 
\\
We stress that the
rare decays $K^+\to\pi^+\nu\bar\nu$ and $K_L\to\pi^0\nu\bar\nu$
are excellent probes of flavor physics. They are not only clean
measures of CKM parameters in their own right, but in addition
complement CP violation studies in $B$ decays due to, in general, 
different sensitivity to new physics and entirely different
experimental systematics.
In particular we emphasize the unique role that can be played by
$K_L\to\pi^0\nu\bar\nu$.
This decay probes directly and unambiguously the nature of CP
violation. Its branching fraction is one of the best measures
of CKM parameters. Especially ${\rm Im}\lambda_t$ and the
Jarlskog parameter $J_{CP}$ can be determined from a $\pm 10\%$
measurement of $B(K_L\to\pi^0\nu\bar\nu)$ with a precision that
can not even be achieved from CP violation studies in $B$ decays
in the LHC era. Of course, the detection of $K_L\to\pi^0\nu\bar\nu$
is experimentally very challenging, but it is not unrealistic.
The current upper limit on the branching fraction is
$5.8\cdot 10^{-5}$ \cite{WEA}. Possibilities for future
experiments have been discussed in \cite{KAMI,ISS}. Recently, a very
interesting proposal has been made, aiming at a $\sim 10\%$
measurement of $B(K_L\to\pi^0\nu\bar\nu)$ at the Brookhaven
Alternating Gradient Synchrotron (AGS) by the year 2000 \cite{DBLL}.
These developments are rather encouraging. The theoretical
motivation clearly warrants all efforts necessary to reach this goal.
It can be expected that further progress will also be achieved for
other quantities, like CP asymmetries in $B$ decays at the 
LHC \cite{LHCB} and $B(K^+\to\pi^+\nu\bar\nu)$ \cite{CCTR,SK},
where the current upper limit of $2.4\cdot 10^{-9}$ \cite{SA} 
is already rather close to the Standard Model expectation of 
$(1.0\pm 0.4)\cdot 10^{-10}$ \cite{BBL}.
\\
The combined use of all available processes will then improve
considerably our understanding of quark mixing and
lead to new, and possibly unexpected, insights into this important
area of high energy physics.

\vspace*{0.7cm}

\noindent {\bf Acknowledgements.} 
We thank Douglas Bryman, Isard Dunietz and Laurence Littenberg
for helpful discussions and suggestions.
This work has been supported by the German Bundesministerium 
f\"ur Bildung und Forschung under contract 06 TM 743 and
by the DFG project Li 519/2-1.
Fermilab is operated by Universities Research Association, Inc.,
under contract DE-AC02-76CHO3000 with the United States Department
of Energy.


\vfill\eject


\begin{thebibliography}{99}
\bibitem{KM}
M. Kobayashi and T. Maskawa, Prog. Theor. Phys. {\bf 49},
652 (1973)
\bibitem{PDG}
R.M. Barnett et al., Particle Data Group, Phys. Rev. {\bf D54}, 
1 (1996)
\bibitem{WO}
L. Wolfenstein, Phys. Rev. Lett. {\bf 51}, 1945 (1983)
\bibitem{BLO}
A.J. Buras, M.E. Lautenbacher and G. Ostermaier,
Phys. Rev. {\bf D50}, 3433 (1994)
\bibitem{BBL}
G. Buchalla, A.J. Buras and M.E. Lautenbacher,
"Weak Decays Beyond Leading Logarithms", FERMILAB-PUB-95/305-T,
{\it to appear in\/} Rev. Mod. Phys.
\bibitem{BB2}
G. Buchalla and A.J. Buras, Nucl. Phys. {\bf B400}, 225 (1993)
\bibitem{BB3}
G. Buchalla and A.J. Buras, Nucl. Phys. {\bf B412}, 106 (1994)
\bibitem{RS}
D. Rein and L.M. Sehgal, Phys. Rev. {\bf D39}, 3325 (1989)
\bibitem{HL}
J.S. Hagelin and L.S. Littenberg, 
Prog. Part. Nucl. Phys. {\bf 23}, 1 (1989)
\bibitem{LW}
M. Lu and M. Wise, Phys. Lett. {\bf B324}, 461 (1994)
\bibitem{FAJ}
S. Fajfer, HU-SEFT-R-1996-05, hep-ph/9602322
\bibitem{MP}
W. Marciano and Z. Parsa, Phys. Rev. {\bf D53}, R1 (1996)
\bibitem{GRO1}
M. Gronau, Phys. Rev. Lett. {\bf 63}, 1451 (1989)
\bibitem{GRO2}
M. Gronau, Phys. Lett. {\bf B300}, 163 (1993)
\bibitem{KPW}
G. Kramer, W.F. Palmer and Y.L. Wu, DESY 95-246, hep-ph/9512341
\bibitem{GL}
M. Gronau and D. London, Phys. Rev. Lett. {\bf 65}, 3381 (1990);
Phys. Lett. {\bf B253}, 483 (1991)
\bibitem{NQ}
Y. Nir and H. Quinn, Phys. Rev. {\bf D42}, 1473 (1990);
Phys. Rev. Lett. {\bf 67}, 141 (1991)
\bibitem{KP}
G. Kramer and W.F. Palmer, Phys. Rev. {\bf D52}, 6411 (1995)
\bibitem{RF}
R. Fleischer, University of Karlsruhe preprint TTP-96-25,
hep-ph/9606469
\bibitem{B94}
A.J. Buras, Phys. Lett. {\bf B333}, 476 (1994)
\bibitem{BH}
A.J. Buras and M.K. Harlander, A Top Quark Story, {\it in}
Heavy Flavors, eds. A.J. Buras and M. Lindner, World Scientific
(1992), p. 58
\bibitem{BB4}
G. Buchalla and A.J. Buras, Phys. Lett. {\bf B333}, 221 (1994)
\bibitem{CJ}
C. Jarlskog, Phys. Rev. Lett. {\bf 55}, 1039 (1985);
Z. Phys. {\bf C29}, 491 (1985)
\bibitem{WEA}
M. Weaver et al., Phys. Rev. Lett. {\bf 72}, 3758 (1994)
\bibitem{KAMI}
K. Arisaka et al., KAMI conceptual design report, FNAL, June 1991
\bibitem{ISS}
T. Inagaki, T. Sato and T. Shinkawa, Experiment to search for the
decay $K_L\to\pi^0\nu\bar\nu$ at KEK 12 GeV proton synchrotron,
30 Nov. 1991
\bibitem{DBLL}
D. Bryman and L. Littenberg, private communications
\bibitem{LHCB}
LHC-B, Letter of Intent, CERN/LHCC 95-5
\bibitem{CCTR}
P. Cooper, M. Crisler, B. Tschirhart and J. Ritchie
(CKM collaboration), 
EOI for measuring $B(K^+\to\pi^+\nu\bar\nu)$ at the Main Injector,
Fermilab EOI 14, 1996
\bibitem{SK}
S. Kettell, talk presented at the AGS-2000 workshop, Brookhaven,
May 1996
\bibitem{SA}
S. Adler et al., Phys. Rev. Lett. {\bf 76}, 1421 (1996)

 
\end{thebibliography}
\end{document}